\documentclass[preprint,aps,nofootinbib,showpacs]{revtex4}
\usepackage{graphics}
\usepackage{epsfig}

\draft

\begin{document}

\title{Diffusion in stochastic sandpiles}

\author{S. D. da Cunha$^{\dagger}$, Ronaldo R. Vidigal$^{\ddagger}$,
L. R. da Silva$^{\dagger}$, and Ronald Dickman$^{\ddagger,*,\S}$}
\address{
$^{\dagger}$Departamento de F\'{\i}sica Te\'orica e Experimental,
Universidade Federal do Rio Grande do Norte, Campus Universit\'ario,
National Institute of Science and Technology for Complex Systems,
59072-970 Natal, Rio Grande do Norte, Brazil\\
$^{\ddagger}$Departamento de F\'{\i}sica, ICEx, Universidade Federal de Minas
Gerais,
Caixa Postal 702, 30161-970 Belo Horizonte, Minas Gerais,
Brazil\\
$^*$National Institute of Science and Technology for Complex Systems,
Caixa Postal 702, 30161-970 Belo Horizonte, Minas Gerais,
Brazil
}
\date{\today}

\begin{abstract}
We study diffusion of particles in large-scale simulations of one-dimensional
stochastic sandpiles, in both the restricted and unrestricted versions.
The results indicate that the diffusion constant scales in the same manner as the
activity density, so that it represents an alternative definition of an order parameter.
The critical behavior of the unrestricted sandpile is very similar to that of its
restricted counterpart, including the fact that a data collapse of the order parameter as a function of
the particle density is only possible over a very narrow interval near the critical point.
We also develop a series expansion, in inverse powers of the density.
for the collective diffusion coefficient in a variant
of the stochastic sandpile in which the toppling rate at a site with $n$ particles is $n(n-1)$,
and compare the theoretical prediction with simulation results.

\vspace{2em}

\noindent$^\S$email: dickman@fisica.ufmg.br
\end{abstract}


\date{\today}
\maketitle

\section{Introduction}

Sandpile models are the prime example of self-organized criticality
(SOC) \cite{btw,dhar99}, or scale-invariance in the apparent absence
of control parameters \cite{ggrin}.  In sandpiles, SOC arises via
a control mechanism that forces the system, which possesses
an absorbing-state phase transition, to its critical point
\cite{bjp,granada}. SOC in a slowly-driven sandpile
corresponds to an absorbing-state phase transition in
the {\it conserved} sandpile, which has
the same local dynamics, but a fixed number of particles
\cite{bjp,tb88,pmb96,vz,dvz,vdmz}.
Conserved sandpiles are
characterized by a nonconserved order parameter (the activity density)
which is coupled to a conserved field that does not evolve in regions
devoid of activity \cite{rossi}.
This class, known as conserved directed percolation (CDP),
is distinct from that of standard directed percolation
\cite{ramasco}.

In recent years considerable progress has been made in
characterizing the critical properties of conserved stochastic
sandpiles, although no complete, reliable theory is yet at hand.  As
is often the case in critical phenomena, theoretical understanding
of scaling and universality rests on the analysis of a continuum
field theory or Langevin equation (a nonlinear stochastic partial
differential equation) that reproduces the phase diagram and
captures the fundamental symmetries and conservation laws of the
system. Important steps in this direction are the recent numerical
studies of a Langevin equation \cite{ramasco,dornic} for CDP.
The critical exponent values reported in Ref. \cite{ramasco} are
in good agreement with those found in simulations of conserved
lattice gas (CLG) models \cite{pastor,kockelkoren}, which exhibit
the same symmetries and conservation laws as stochastic sandpiles.
The Langevin equation exponents are also consistent with the best
available estimates for stochastic sandpiles in two dimensions
\cite{ramasco}.
There is now good evidence that the one-dimensional stochastic sandpile
belongs to the CDP universality class \cite{mnrst2,munoz08}.

In this work we focus on an aspect of sandpiles that has received relatively little attention:
diffusion.  Since the dynamics in these models involves hopping of particles
between neighboring sites, one expects the
particle diffusion constant $D$ to follow a scaling behavior similar to that of the
usual order parameter $\rho$.  (A site is active if it
bears two or more particles.)
Here $D$ is
defined via the relation $\langle (\Delta x)^2 \rangle = 2Dt$,
where $\Delta x$ is the particle displacement.   We determine the scaling
properties of the diffusion constant in extensive Monte Carlo simulations.

Theoretical studies of the particle diffusion coefficient are hampered by the fact that
positions of specific particles are not accessible in the usual stochastic description;
instead the master equation describes the evolution of the probability distribution on
the set of occupation numbers $\{n_i\}$.  It is however possible to determine the
{\it collective} diffusion coefficient $D_c$ by studying how a density perturbation
$\Delta p \propto e^{ikx}$ relaxes.  $D_c$ is related to the relaxation time, $\tau_k$,
of this mode via $\tau_k = 1/(D_c k^2)$, in the small-$k$ limit. Using the
path-integral based perturbation theory developed in \cite{manpert}, we calculate the first
three terms in the expansion of $D_c$ in inverse powers of density $p$, for the stochastic
sandpile in which the toppling rate is $n(n-1)$.


The balance of this paper is organized as follows. In Sec. II we define
the three models of interest. Sec III reports simulation results on $D$ and $\rho$
for two of these models. In Sec IV we develop a series expansion for
the collective diffusion coefficient in a third model, and compare the predictions with
simulation results.  We close in Sec. V with a summary and discussion.

\section{Models}

We study three versions of the one-dimensional conserved stochastic sandpile,
related to Manna's model \cite{manna}.
In these systems the configuration is defined by the set of occupation
variables $n_1,...,n_L$, giving the number of particles residing at each site
on a ring of size $L$.
All three versions are continuous-time Markov processes, in which transitions
involve the ``toppling" of an active site, i.e., one with $n_i \geq 2$.
The particular features distinguishing the three models are as follows.

{\it Basic unrestricted model (I)} \cite{manna1d}.  Each active site has a rate of unity to topple.
When site $i$ topples, two particles are transferred from this site to its
neighboring sites ($i-1$ and/or $i+1$, with periodic boundaries).  The two particles
jump independently; they jump to the left or to the
right with equal probabilities.

{\it Restricted model (II)}  \cite{mnrst,mnrst2}.  The dynamics is that of model I
except that no site may have more than two particles.  If, when a site topples,
a particle attempts to jump to a site bearing two particles, it returns to the toppling site.

{\it Modified unrestricted model (III).}  \cite{manpert}. The dynamics is that of model I
except that the rate at which a site topples is given by $n_i (n_i-1)$.

While somewhat less convenient for simulation,
model III is better suited to operator-based theoretical approaches.  Model II features a
smaller set of states, rendering it more convenient for analysis via
cluster approximations \cite{mancam}.  There is clear numerical evidence that model II belongs
to the CDP universality class \cite{mnrst2}; models I and III share the same symmetries and
conserved quantities as model II and so are expected to belong to the CDP class as well.

In conserved sandpiles, the particle density $p$ serves as a temperaturelike
control parameter.  Below a certain critical value, $p_c$, the system eventually
falls into an absorbing configuration (i.e., one devoid of active sites), while for
$p>p_c$, activity continues indefinitely, in the infinite-size limit.  The
order parameter associated with this absorbing-state phase transition is the
activity density $\rho$, given by the stationary mean fraction of active sites in models I and II,
and the stationary average $\langle n_i(n_i-1) \rangle$ in model III.  Numerical studies
strongly support a continuous transition at $p_c$; best estimates for the critical
density $p_c$ are 0.9488, 0.92978, and 0.9493, in models I, II, and II, respectively.
(Note that for $p > 1$ active sites always exist; $p_c$, however, is strictly less than
unity.)
\vspace{1em}

\begin{figure}[ht]
\vspace{2em}
\begin{center}
\epsfxsize=80mm
\epsffile{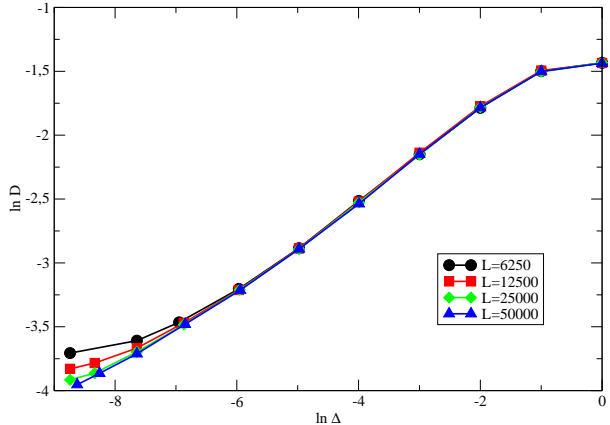}
\caption{(Color online) Asymptotic diffusion rate versus $\Delta = p - p_c$ in model I,
system sizes as indicated.  Error bars are smaller than the symbols.}
\label{D_nao_restrito}
\end{center}
\end{figure}
\vspace{3em}

\begin{figure}[ht]
\begin{center}
\epsfxsize=85mm
\epsffile{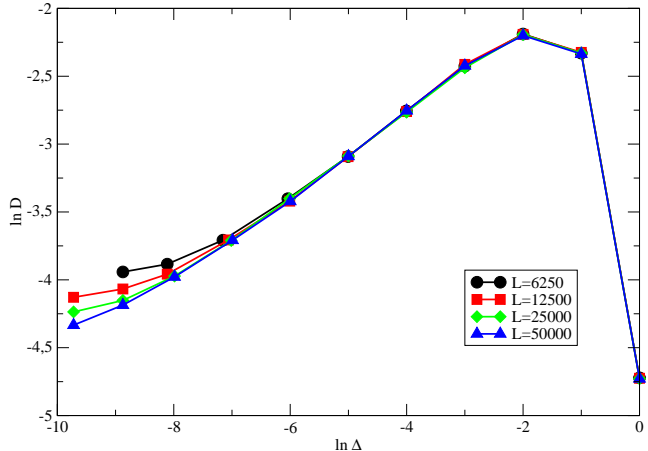}
\caption{(Color online) Asymptotic diffusion rate versus $\Delta$ in model II.}
\label{D_restrito}
\end{center}
\end{figure}

\section{Simulation results}

In this section we report simulation results for the particle diffusion rate $D$
and the activity density $\rho$ in models I and II.  For each particle $j$,
let $\Delta_j (t) = h_j^+ (t) - h_j^- (t)$, where $h_j^+ (t)$ and $h_j^- (t)$
are the numbers of hops taken by particle in the positive (respectively, negative) directions up to time $t$.
Then the particle diffusion rate is defined through the relation

\begin{equation}
\langle [\Delta_j (t)]^2 \rangle = 2 D t
\label{deffD}
\end{equation}

\noindent where the average is over all particles and (in principle) all histories of the system
of $N = pL$ particles on $L$ sites, starting from a given initial configuration or class of
configurations.  (In practice we generate initial configurations by adding particles
randomly to the system, with the prohibition, in the case of model II, of triple or higher
occupancy.  We average over a set of $N_s$ realizations of the process.)
Note that the diffusion rate
$D$ defined above depends in general on the time $t$ as well as on $L$ and $p$.
Determination of $D$ in simulations requires that we store the particle displacements,
which is not necessary if we merely wish to study the activity density.  Since $\Delta(t)$
refers to a given history of the process, the quasistationary simulation method
used in \cite{mnrst2} is not applicable here.

Each time particle $j$ hops, $\Delta_j$ changes by $\pm 1$, so that
$\langle [\Delta_j (t)]^2 \rangle = \langle h_j^+ (t) + h_j^- (t) \rangle \equiv \langle h_j (t) \rangle$,
i.e., the mean number
of jumps up to time $t$.  Particle $j$ must be at an active site in order to jump,
but $\langle h_j (t) \rangle$ is not simply equal to $2 \rho$, as would be the case
if particle $j$ were always to jump each time the site it resides at topples.  In model I, for example,
the probability of a given particle jumping is $2/n$, if it is one of $n$ particles at the toppling
site.  In model II, the particle always tries to hop when its host site topples, by it
may be unable to move to the target site.  Finally, in model III a particle at a site
with occupation $n$
hops at a rate of $2(n-1)$, so that the hopping rate actually
grows with the occupation number $n$.
Since the occupancies of nearby sites are correlated, the waiting times between successive displacements
of a given particle are not independent.  For these reasons, the relation between the hopping
rate and the activity density involves subtle effects, different in each of the three models
studied.  It is nevertheless reasonable to expect that, as $p \to p_c$, the scaling behavior
of the diffusion rate will parallel that of the activity density.  In particular, we might
expect $\rho$ and $D$ to be governed by the same set of critical exponents at the transition.

We simulate models I and II on rings of $L=$ 6250, 12500, 25000, and 50000 sites, using eight independent
realizations for the smallest size, six for $L=12500$, and four for the two largest sizes.
The studies are run for 10$^6$ to $6 \times 10^9$ time units, with the longest simulation times
near the critical point.
Each particle is assigned a label so that its cumulative dislocation $\Delta_j$ can be
followed during the evolution.
In model II, when two particles attempt to jump to the same site, and this site is singly occupied,
one of the two particles is chosen at random to move to the target site, while the other remains
where it is.

We monitor the diffusion rate $D(t)$, defined in Eq (\ref{deffD}), and confirm that it
approaches a stationary value at long times.  Figs. \ref{D_nao_restrito}
and \ref{D_restrito} show the stationary value, $D$, as a function of $\Delta = p - p_c$, for
models I and II, respectively.  Several aspects of these results are worth commenting on.
First, for the sizes considered here, $D$ is apparently well converged to its limiting
($L \to \infty$) value for $\Delta \geq 0.0025 $.  Second, even for values of $\Delta$
such that the diffusion rate has converged,
the slope of $D(\Delta)$ on logarithmic
scales changes appreciably with $\Delta$, making a reliable estimate
of $\beta$ difficult.  Finally, in model II, $D$ drops sharply as $p$ approaches 2: due to the
height restriction, most particles cannot move.

\begin{figure}[t]
\begin{center}
\epsfxsize=85mm
\epsffile{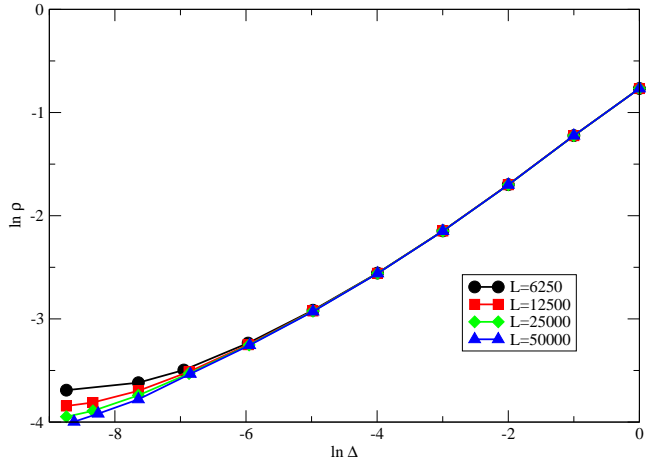}
\caption{(Color online) Stationary activity density $\rho$ versus $\Delta$ in model I.}
\label{rho_nao_restrito}
\end{center}
\end{figure}

\begin{figure}[t]
\begin{center}
\epsfxsize=85mm
\epsffile{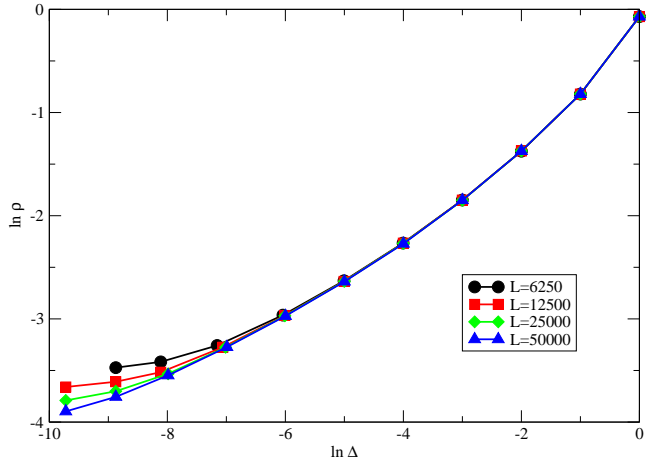}
\caption{(Color online) Stationary activity versus $\Delta$ in model II.}
\label{rho_restrito}
\end{center}
\end{figure}

\begin{figure}[ht]
\begin{center}
\epsfxsize=120mm
\epsffile{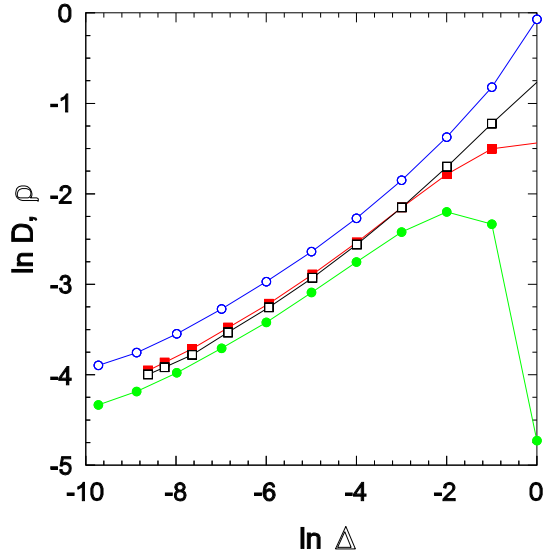}
\caption{(Color online) Stationary activity (open  symbols) and asymptotic diffusion rate
(filled symbols) versus $\Delta$ in models I (squares) and II (circles),
system size $L=50\,000$.}
\label{dandr50}
\end{center}
\end{figure}

The corresponding results for the activity density $\rho$ are shown in Figs. \ref{rho_nao_restrito}
and \ref{rho_restrito}, respectively.  The behaviors of $D$ and $\rho$ in both models appear quite similar,
an impression that is confirmed in Fig. \ref{dandr50}, which compares both quantities (in both models),
for the
largest system studied.  Near the transition, $\rho$ and $D$ are virtually identical in the
unrestricted model, while in model II they appear to be proportional.
It is evident that neither $\rho$ nor $D$ can be characterized as following a simple power law,
an observation already made for the order parameter in model II in \cite{mnrst2}.

\begin{figure}[ht]
\begin{center}
\epsfxsize=85mm
\epsffile{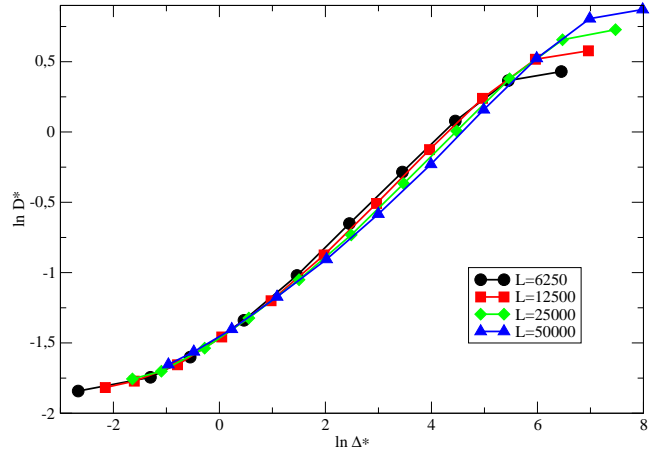}
\caption{(Color online) Scaled diffusion rate $D^* = L^{\beta/\nu_\perp} D $ versus scaled distance
from critical point $\Delta^* = L^{1/\nu_\perp} \Delta$ in model I.}
\label{DNRcol}
\end{center}
\end{figure}

Note that for the system sizes studied here, there is no discernable finite-size effect for
$p - p_c \geq 0.0025$.  Since $D(\Delta)$ and $\rho(\Delta)$ do not follow simple power laws in
this regime, there is no possibility of maintaining the data collapse under the usual kind
of FSS scaling plot, that is, of $\rho^* = L^{\beta/\nu_\perp} \rho $ versus $\Delta^* =
L^{1/\nu_\perp} \Delta$.
As noted in \cite{mnrst2}, a data collapse can only be achieved in the regime very near the
critical point (i.e., $\Delta \leq 0.0025$); and example of a data collapse for the diffusion rate
data is shown in Fig. \ref{DNRcol}.  In model I the data collapse is best using exponent values
$\beta=0.289$ and $\nu_\perp=1.35$; the corresponding values in model II are
$\beta=0.285$ and $\nu_\perp=1.355$.  These values are consistent with those reported in \cite{mnrst2}:
$\beta=0.289(12)$ and $\nu_\perp=1.355(18)$.
We may therefore affirm, with a high degree of confidence, that models I and II belong to the same
universality class, and that the diffusion rate and the order parameter exhibit the same critical
scaling properties.

\begin{figure}[ht]
\begin{center}
\epsfxsize=100mm
\epsffile{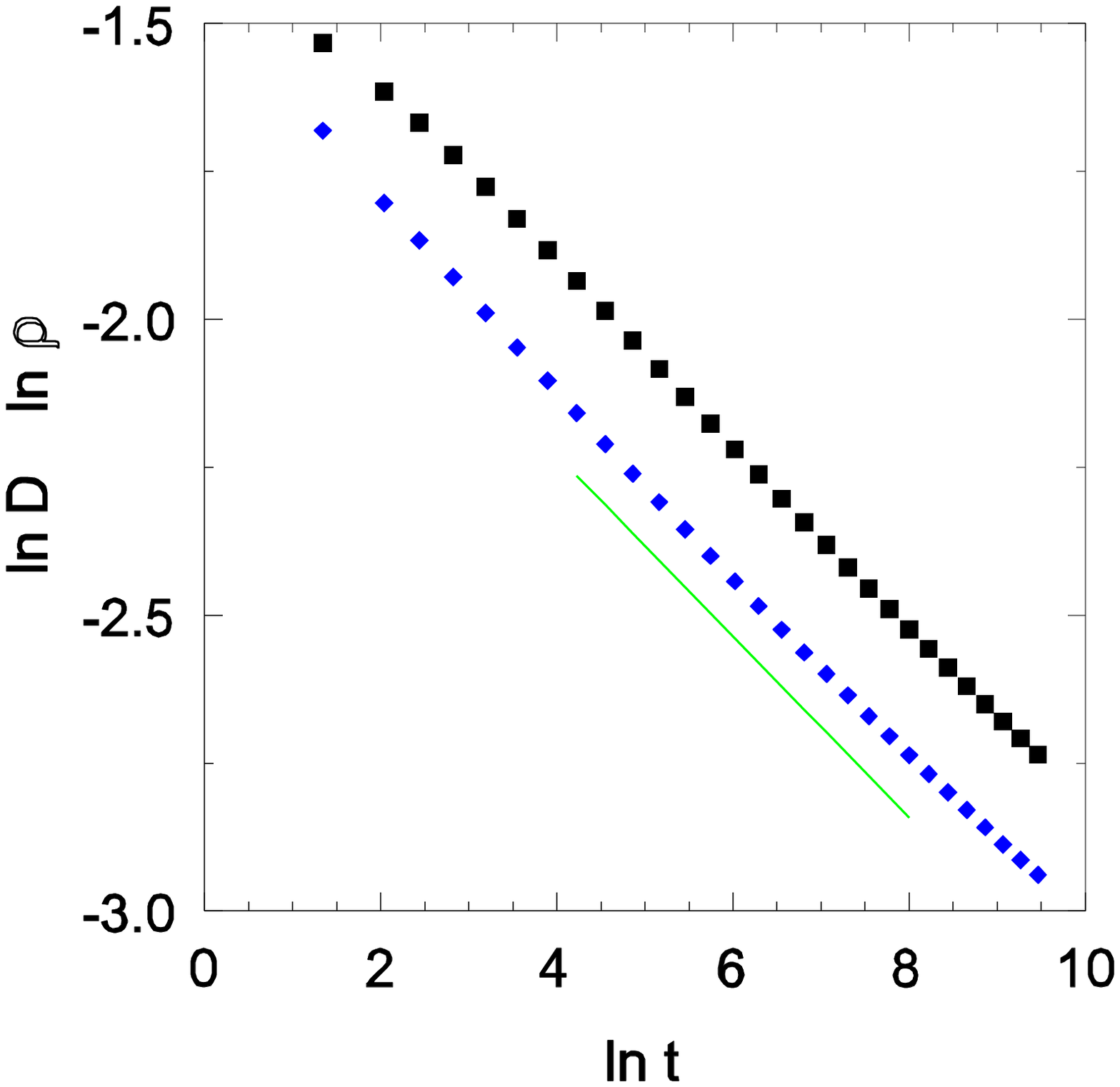}
\caption{(Color online) Initial decay of activity (upper) and diffusion rate
(lower) at criticality in model I;
system size $L=50\,000$.  The slope of the straight line is -0.153}
\label{rdvst}
\end{center}
\end{figure}

In the paradigmatic examples of absorbing-state phase transitions, such as the contact
process \cite{torre,marro}, starting from a spatially homogeneous initial distribution,
the order parameter exhibits an initial power-law decay,
$\rho \sim t^{-\delta}$, at the critical point, before saturating at a quasistationary
value.  (Note that the power-law portion of the evolution is independent of system size.)
It is of interest to know whether the order parameter and the diffusion rate exhibit similar behavior
in the stochastic sandpile.  Our results (Fig. \ref{rdvst}) for $L=50 \, 000$
show $\rho$ and $D$ decaying with an exponent $\delta = 0.153(5)$.

\begin{figure}[ht]
\begin{center}
\epsfxsize=85mm
\epsffile{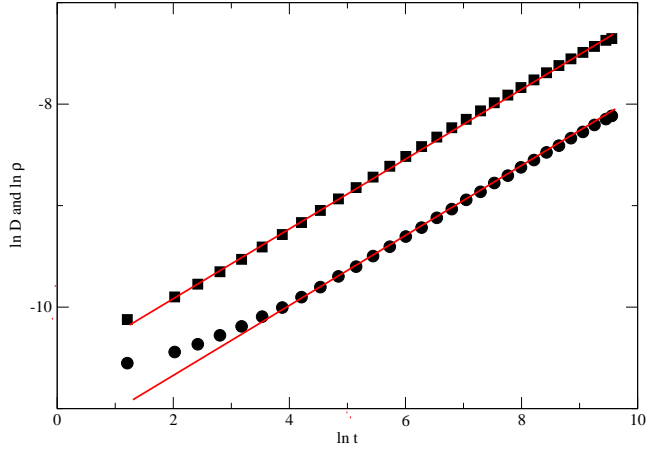}
\caption{(Color online) Initial growth of activity (squares) and diffusion rate
(circles) at criticality in model II, starting with a single active site;
system size $L=50\,000$.}
\label{thR50k}

\end{center}
\end{figure}

We also perform simulations of the spread of activity with time.  In this case, a single site is
given two particles initially, while the remaining $N-2$ particles are distributed at random,
one per site, over the rest of the lattice.  In the contact process at criticality, starting with a single
active site, the number of active sites grows as $n(t) \sim t^\eta$ \cite{torre,marro}.  In the present case
we find that both $D$ and $\rho$ follow an approximate power law with an exponent $\eta = 0.34(1)$, as shown in
Fig. \ref{thR50k}.  The spreading exponents $\delta$, $\eta$, and $z_{sp}$ are expected to satisfy
the hyperscaling relation $4 \delta + 2 \eta = d z_{sp}$, which, using $z_{sp} = 2/z$, with $z$
the usual dynamic exponent, can be written as $\eta = 1/z - 2 \delta$.  Using the value of $\delta$ cited
above, and $z = 1.50(4)$ from Ref. \cite{mnrst2}, this yields $\eta = 0.36(3)$, which is
consistent with our numerical estimate.

\section{Collective diffusion coefficient: theory and simulation}

In this section we apply the operator formalism and perturbation theory
derived in \cite{manpert} to the evaluation of the collective diffusion
coefficient $D_c$ of model III on a ring of $N$ sites.
We begin by writing the master equation for the process
in the form

\begin{equation}
\frac{d |\Psi \rangle}{dt} = L |\Psi \rangle ,
\end{equation}
where
\begin{equation}
|\Psi \rangle = \sum_{\{n\}} p(\{n\},t) |\{n\} \rangle
\end{equation}
is the probability distribution.
Here $p(\{n\},t)$ is the probability of configuration
$\{n\}$, and the state $|\{n\} \rangle$
is a direct product of states $|n_j \rangle$,
representing exactly $n_j$ particles at site $j$.
These states are normalized so: $\langle n' | n \rangle = n! \delta_{n,n'}$.

Defining
creation and annihilation operators via the relations,

\begin{equation}
a_i |n_i\rangle = n_i |n_i\!-\!1\rangle
\end{equation}
and
\begin{equation}
\pi_i |n_i\rangle = |n_i\!+\!1\rangle ,
\end{equation}

\noindent so that $[a_i,\pi_j] = \delta_{ij}$,
the evolution operator for the one-dimensional stochastic sandpile is
\begin{equation}
L = \sum_i \left[ \frac{1}{4} (\pi_{i-1} + \pi_{i+1})^2
- \pi_i^2 \right] a_i^2 .
\label{evop}
\end{equation}

Since the system is translation-invariant it is convenient to
introduce the discrete Fourier transform via

\begin{equation}
a_k = \sum_j e^{-ijk} a_j \;,
\end{equation}
with inverse
\begin{equation}
a_j = \frac{1}{N} \sum_k e^{ijk} a_k \;,
\end{equation}
(and similarly for other variables),
where the allowed values of the wavevector are:

\begin{equation}
k = -\pi, \; -\pi \!+\! \frac{2\pi}{N}, ...
-\frac{2\pi}{N}, \; 0, \;
\frac{2\pi}{N}, ..., \pi \!-\! \frac{2\pi}{N} \;.
\label{brill}
\end{equation}

\noindent (To avoid heavy notation, we
indicate the Fourier transform by the subscript $k$;
the subscript $j$ denotes the corresponding variable on the lattice.)
In the Fourier representation, the evolution operator takes the form

\begin{equation}
L = -N^{-3} \sum_{k_1,k_2,k_3} \omega_{k_1,k_2} \pi_{k_1} \pi_{k_2} a_{k_3}
a_{-k_1-k_2-k_3},
\label{evopk}
\end{equation}

\noindent where $\omega_{k_1, k_2} = 1-\cos k_1 \cos k_2$.
As explained in Ref. \cite{manpert},
the evolution operator may be rewritten as
\begin{equation}
L = L_0 + L_1
\end{equation}
\noindent with
\begin{equation}
L_0 = - N^{-1} \sum_{k \neq 0} \gamma_k \pi_{-k} a_k,
\end{equation}
\noindent and
\begin{eqnarray}
\label{Eq:LI}
L_1 & = & -N^{-3} \sum_{k_{1},k_{2},k_{3} \neq 0}
\omega_{k_{3},-k_{1}-k_{2}-k_{3}} \pi_{k_{3}}
\pi_{-k_{1}-k_{2}-k_{3}}  a_{k_{1}} a_{k_{2}}
 \nonumber \\
& - &  2pN^{-2}\sum_{k_1,k_2 \neq 0}
\omega_{k_2,-k_1-k_2}  \pi_{k_2}
\pi_{-k_{1}-k_2} a_{k_1}
\nonumber \\
& -&  2N^{-2}  \sum_{k_{1},k_{2} \neq 0}
 \omega_{-k_{1}-k_{2},0} \pi_{-k_{1}-k_{2}} a_{k_{1}} a_{k_{2}}
- p^{2}N^{-1} \sum_{k \neq 0} \omega_{k,-k} \pi_k
\pi_{-k},
\end{eqnarray}
\noindent where

\begin{equation}
\gamma_k = 4 p \, \omega_{k,0} = 4p(1 - \cos k).
\end{equation}
This transformation is based on the observation that, due to
particle conservation, the operator $N^{-1} \sum_j \pi_j a_j $ may be
equated to the particle density $p$.
In Eq. (\ref{Eq:LI}), it is understood that none of the wavevectors associated with
the operators $a$ and $\pi$ may be zero.

Let $P_n = e^{-p} p^n/n! $ denote the Poisson distribution with intensity $p$, and define
$|P\rangle_i = \sum_{n} P_n |n \rangle_i $ as the Poisson-distributed state at site $i$.
Then the uniform product-Poisson distribution is $|P\rangle \equiv \otimes_i |P\rangle_i$.
The latter is an eigenstate of the diffusion operator with eigenvalue zero,
i.e., ${\cal D} |P\rangle = 0$, where

\begin{equation}
{\cal D} = \frac{1}{2} \sum_j \left[ \pi_{j-1} - 2 \pi_j + \pi_{j+1} \right] a_j
         = - N^{-1} \sum_k \omega_{k,0} \pi_{-k} a_k,
\label{diffop}
\end{equation}

\noindent represents nearest-neighbor hopping at unit rate.
(Note that $L_0 = 4p {\cal D}$.)

\vspace{3em}

\begin{figure}[ht]
\begin{center}
\epsfxsize=120mm
\epsffile{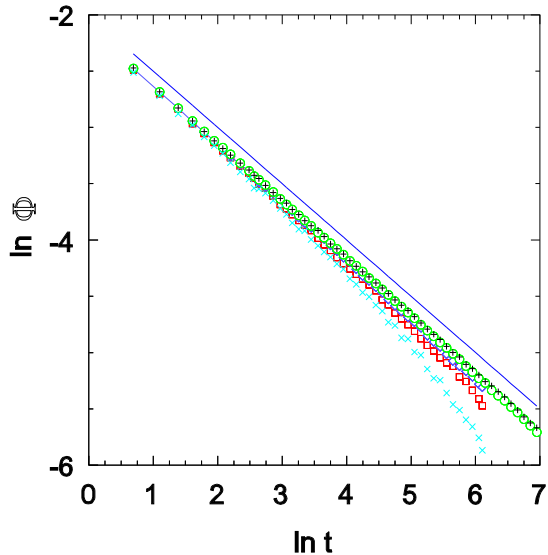}
\caption{(Color online) Projection $\Phi(t)$ in simulations of model III.  System sizes (lower to upper)
$N=400$, 800,..., 6400, particle density $p=3.0$.  The slope of the straight line is $-1/2$.}
\label{ftp3}
\end{center}
\end{figure}

To study collective diffusion, we consider an initial condition in which the uniform
Poisson-product is weakly perturbed by a density modulation with wavevector $k$:

\begin{equation}
|\Psi(0)\rangle = N^{-1} \pi_k |P \rangle
\label{Psizero}
\end{equation}

\noindent Introducing the notation,

\begin{equation}
\langle \; | \equiv \sum_{\{ n \}} \prod_j \frac{1}{n_j!} \langle n_j |
\label{normbra}
\end{equation}

\noindent for the projection onto all possible configurations, the mean number of
particles at site $j$ is given by

\begin{equation}
\phi_j(t) = \langle n_j(t) \rangle = \langle \; | a_j | \Psi(t) \rangle
\label{phijt}
\end{equation}

\noindent or equivalently, in the Fourier representation,

\begin{equation}
\phi_k(t) = \langle \; | a_k | \Psi(t) \rangle.
\label{phikt}
\end{equation}

\noindent Note that for the initial distribution of Eq. \ref{Psizero}, with $k \neq 0$,
we have

\begin{equation}
N^{-1} \langle \; | a_q \pi_k | P \rangle = \delta_{q,-k}
\end{equation}

\noindent where we used the relations $[a_q, \pi_k] = N \delta_{q, -k}$ and
$\langle \; | \pi_k | P \rangle = N\delta_{k,0}$.
Thus $\phi_{-k}(t)$ represents the amplitude, at time $t$, of the
density perturbation created at time zero.

We assume that for long times
and long wavelengths the mean density $\phi_j$ satisfies the diffusion equation
$\partial \phi_j / \partial t = D_c \Delta^2 \phi_j$ with $\Delta^2$ the discrete Laplacian,
leading, in the small-$k$ limit, to $\phi_k (t) \simeq \phi_k (0) \exp[-D_c k^2 t]$.
Letting $\phi_k (z)$ denote the Laplace transform, we have, in the small-$z$ limit,
$\phi_k (z) \simeq 1/(z + D_c k^2)$, so that,

\begin{equation}
D_c = \lim_{k, z \to 0} \, \frac{1}{k^2 \phi_k (z)}.
\label{Dc1}
\end{equation}
\vspace{1em}

\begin{figure}[ht]
\begin{center}
\epsfxsize=120mm
\epsffile{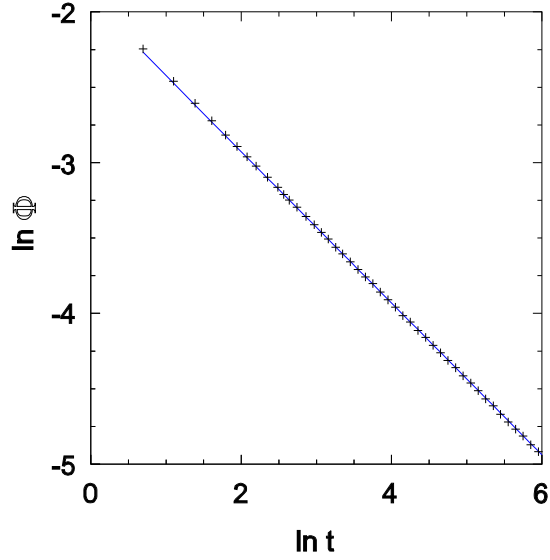}
\caption{(Color online) Projection $\Phi(t)$ in simulations of model III
with $N=12800$ and $p=2.0$. The solid line is given by Eq.~(\ref{Fhit}) with $D_c=3.76$.}
\label{ft77a}
\end{center}
\end{figure}


Laplace transforming the formal solution of the master equation we find

\begin{equation}
\phi_{-k} = N^{-1} \langle \; | a_{-k} \, \frac{1}{z-L} \, \pi_k |P \rangle
\label{phikz}
\end{equation}

\noindent We may develop the solution in a series in powers of $1/p$ using the
operator identity

\begin{equation}
\frac{1}{z-L_0 -L_1} = \frac{1}{z-L_0} + \frac{1}{z-L_0} L_1 \frac{1}{z-L_0}
+ \frac{1}{z-L_0} L_1 \frac{1}{z-L_0} L_1 \frac{1}{z-L_0} + \cdots
\label{identL}
\end{equation}

\noindent Evaluation of the contributions to this series is facilitated by use of the
identities $L_0 | P \rangle = 0$, $a_k | P \rangle = N p \delta_{k,0} |P \rangle$,
and,

\begin{equation}
\langle \; | a_q \pi_k  | P \rangle = N  \delta_{q,-k} + N^2 p \delta_{q, 0} \delta_{ k, 0}.
\label{id1}
\end{equation}

\noindent Note also that $ L_0 \pi_k |p\rangle = - \gamma_k \pi_k |P \rangle$, and
in general,

\begin{equation}
L_0 \, \pi_{k_1} \cdots \pi_{k_n}  | P \rangle = - {\cal S} \, \pi_{k_1} \cdots \pi_{k_n}  | P \rangle.
\label{id2}
\end{equation}

\noindent where ${\cal S} = \gamma_{k_1} + \cdots + \gamma_{k_n}$.  Thus $L_0$ may be inverted on the
space of states of the form $\pi_{k_1} \cdots \pi_{k_n}  | P \rangle$ provided that not all
of the wave vectors are zero: on this space $L_0^{-1}$ is simply $- 1/{\cal S}$ times the identity
operator.

The first term in the expansion of Eq. (\ref{phikz}) is readily evaluated as,

\begin{equation}
\phi_{-k}^{(0)} = N^{-1} \langle \; | a_{-k} \, \frac{1}{z-L_0} \, \pi_k |P \rangle
= \frac{1}{z+ \gamma_k},
\label{phikz0}
\end{equation}
\vspace{1em}

\noindent which gives $\lim_{k, z \to 0} \, k^2 \phi_{-k} (z) = 1/(2p) + {\cal O}(1/p^2)$.
Subsequent terms in the expansion may be evaluated using the diagrammatic perturbation
approach developed in \cite{manpert}.  In this representation each term in $L_1$
corresponds to a vertex, with operators $a_k$ corresponding to lines entering the vertex
at the right, and operators $\pi_k$ to lines leaving at the left.  Each line that leaves
a vertex must be joined (``contracted") with a line entering some other vertex to the left.
The operator $L_1$, Eq. (\ref{Eq:LI}), consists
of four parts or vertices, designated respectively as a crossing (two lines in, two out),
a bifurcation (one in, two out), a conjunction, and a source.  We denote these contributions as
$L_a$, $L_b$, $L_c$ and $L_d$, respectively. There are two diagrams that contribute to
$\phi_{-k}(z)$ at order $1/p^2$. One arises from the term,

\begin{eqnarray}
N^{-1} \langle \; | a_{-k} \frac{1}{z\!-\!L_0} L_c \frac{1}{z\!-\!L_0} L_b \frac{1}{z\!-\!L_0}
\pi_k |P \rangle &=&
\frac{1}{8Np^2} \sum_{q \neq 0} \frac{1 \!-\! \cos q \cos(k\!-\!q)}
{2 \!-\! \cos q - \cos(k\!-\!q)}
\nonumber
\\
&=& \frac{1}{16p^2 (1 \!-\! \cos k)},
\label{diag1}
\end{eqnarray}

\noindent while the second is,

\begin{equation}
N^{-1} \langle \; | a_{-k} \frac{1}{z\!-\!L_0} L_c \frac{1}{z\!-\!L_0} L_c \frac{1}{z\!-\!L_0}
L_d \frac{1}{z\!-\!L_0} \pi_k |P \rangle = - \frac{1}{32p^2 (1 \!-\! \cos k)}.
\label{diag2}
\end{equation}
\vspace{1em}

\noindent At order $1/p^3$ there are 17 diagrams, leading to

\begin{equation}
D_c = \frac{2p}{1 + \frac{1}{8p} + \frac{0.1088899}{p^2} + \cdots}
\label{Dpert}
\end{equation}
\vspace{1em}

In \cite{manpert} the stationary activity density $\rho = \langle n(n-1) \rangle$ was found to grow asymptotically
as $p^2$, with correction in inverse powers of $p$; here we find that $D_c$ grows only linearly with $p$.
For comparison we write the results for $\rho$ and $D_c$ in the form:

\begin{equation}
\rho = p^2 \left[ 1 - \frac{1}{4p} - \frac{0.0492525}{p^2} + \cdots \right]
\label{rhopert}
\end{equation}
\noindent and
\begin{equation}
D_c = 2p \left[ 1 - \frac{1}{8p} - \frac{0.093265}{p^2} + \cdots \right]
\label{Dpert1}
\end{equation}
\vspace{1em}

\noindent These expressions are reliable for $4p \gg 1$ but cannot of course be applied
in the vicinity of the critical density, $p_c \simeq 0.9493$.

\begin{figure}[ht]
\begin{center}
\epsfxsize=120mm
\epsffile{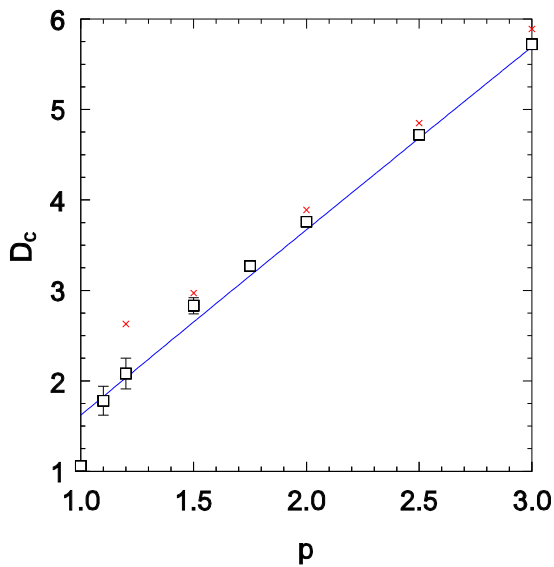}
\caption{(Color online) Collective diffusion constant $D_c$ versus particle density $p$
in model III.  Squares: simulation; line: theory, Eq. (\ref{Dpert}).  The crosses denote simulation
values for the stationary-state collective diffusion constant $D_{c,s}$.
}
\label{mandif}
\end{center}
\end{figure}

\subsection{Comparison with simulation}

We determine the collective diffusion coefficient in model III via analysis
of the {\it projection} of the configuration $\{n(j,t)\}$ on the initial
configuration.  Recalling that $\langle n_j(t) \rangle = p$,
the particle density, we let $f_j(t) \equiv n_j(t) - p$ denote the excess particle
number at site $j$ and time $t$.  Initially, the $\{n_j\}$ are independent, Poisson-distributed
with mean $p$.  Consider now

\begin{equation}
\Phi(t) \equiv \frac{\langle f_j(t) f_j(0) \rangle}{\langle f_j(0)^2 \rangle}
\label{defPhi}
\end{equation}

\noindent where the angular brackets denote an average over sites and over realizations, including
the random initial configuration.  For the set of wavevectors $k$ defined in Eq. (\ref{brill}),
let $\varphi_k (t)$ denote the discrete Fourier transform,

\begin{equation}
\varphi_k (t) = \sum_{j=1}^N f_j (t) e^{ijk}
\end{equation}
\vspace{1em}

In the small-$k$ limit we expect $\varphi_k$ to follow,
\begin{equation}
\varphi_k (t) = \varphi_k (0) e^{-D_c k^2 t}
\end{equation}

\noindent Using the fact that the $f_j(0)$ are independent, zero-mean
random variables with var$[f_j(0)] = p$, it is straightforward to show that

\begin{equation}
\Phi(t) = \frac{1}{N} \sum_k e ^{-D_c k^2 t}
\simeq \frac{1}{2\pi} \int_{-\pi}^\pi e ^{-D_c k^2 t} dk
\label{Phit}
\end{equation}

\noindent For times such that $D_c \pi^2 t \gg 1$ we may extend the limits of integration to
$\pm \infty$, yielding

\begin{equation}
\Phi \simeq \frac{1}{2\pi} \sqrt{\frac{\pi}{D_c t}}
\end{equation}

\noindent Thus if $\Phi (t)$, as determined via simulation, can be fit for large $t$ with an expression
of the form $A/t^{1/2}$, then
$D_c = 1/(4 \pi A^2)$.  In practice, however, a more reliable procedure is to fit the
simulation data to the full lattice expression

\begin{equation}
{\cal F}(t) \equiv \frac{1}{N} \sum_k e ^{-2 D_c [1 - \cos k]  t}
\label{Fhit}
\end{equation}

\noindent which involves the single adjustable parameter $D_c$, and is capable of fitting
the data at short as well as long times, and for various system sizes.

We determine $\Phi(t)$ on rings of $N= 200, 400, 800,..., 25600$ sites, for particle densities
$p$ in the range of 1 to 3.  Fig. \ref{ftp3} shows that as the system size increases, $\Phi (t)$ approaches
a power-law decay with an exponent of 1/2; an example of data fit by Eq. (\ref{Fhit}) is
shown in Fig.~\ref{ft77a}.
The estimates for $D_c$, obtained by fitting ${\cal F}(t)$ to the simulation data, are compared
with the theoretical prediction, Eq.~(\ref{Dpert}), in Fig.~\ref{mandif};  the agreement is quite good
for densities $p \geq 2$.
In Fig.~\ref{mandif},
the relatively large error bars associated with the simulation results for $p=1.1$ and 1.2
reflect the fact that the simulation results for $\Phi(t)$ are less well fit by the
theoretical expression, Eq. (\ref{Fhit}), than for other particle densities.  Curiously,
for $p=1$, despite being nearer the critical point, the fit is again quite good.

A similar analysis can be applied to extract the value of $D_{c,s}$ from simulations in the stationary state.
In this case, we allow the system to relax, so that the configuration at time zero is typical
of the stationary distribution.  Now, however, the $f_j(0)$ are no longer independent, Poisson
distributed variables, and the power spectrum of fluctuations $\langle |\varphi_k(0)|^2 \rangle$ is no longer constant.
We therefore fit the data for $\Phi(t)$ using the expression

\begin{equation}
{\cal F}_s(t) \equiv \frac{ \sum_k \langle |\varphi_k(0)|^2 \rangle e ^{-2 D_{c,s} [1 - \cos k]  t}}
{\sum_k \langle |\varphi_k(0)|^2 \rangle}
\label{Fhits}
\end{equation}
\vspace{.5em}

\noindent with $\langle |\phi_k(0)|^2 \rangle$ determined via simulation.  The resulting stationary
values of $D_{c,s}$ are close to, but slightly greater than, those found using the Poisson initial
distribution (see Fig.~\ref{mandif}).  It is worth noting that in the stationary state the projection
$\Phi(t)$ appears to decay with a power smaller than 1/2 (a typical exponent value is about 0.41).
This does not imply anomalous behavior as the data can again be fit using the hypothesis
$\varphi_k (t) = \varphi_k (0) e^{-D_c k^2 t}$.  For densities $p < 1.2$ however, the simulation data
are not well fit by the function ${\cal F}_s(t)$.  In this regime the Fourier amplitudes
$\langle \varphi_{-k} (t) \varphi_k (0) \rangle$ (calculated in simulations) do not follow a
simple exponential decay.  (The data suggest a crossover to stretched-exponential decay at long times.)
Thus, near the critical
point, we find evidence of anomalous relaxation, as previously noted
in stochastic sandpiles \cite{mannarel}.

\section {Summary}

We study diffusion in stochastic sandpiles.  In the first part of this work we determine
the particle diffusion coefficient in sandpiles in which all active sites share the
same toppling rate.  We find, in both the restricted and unrestricted cases, that
the diffusion constant scales in the same manner as the order parameter (the activity
density).  Our results confirm that the restricted and unrestricted models belong to
the same universality class, and that both models exhibit a finite-size scaling collapse of data
over an unusually narrow region of the control parameter (that is, the particle density $p$).

The second part of this study deals with a sandpile
in which the toppling rate at site $i$ is $n_i (n_i -1)$.  In this case it is possible
to derive a short series for the collective diffusion constant, starting from a
Poisson-product initial state.  The resulting expression compares well with
simulation for densities well above $p_c$.  The collective diffusion constant $D_c$ is
extracted from simulations using the projection of
density fluctuations at time $t$ onto their initial values.
We expect this approach to
be useful in determining $D_c$ in other systems, such as interacting lattice gases.
We defer a detailed investigation of collective diffusion in the critical region
to future work.

\vspace{2em}

\noindent{\bf Acknowledgements}
\vspace{1em}

We thank Alvaro Vianna Novaes de Carvalho Teixeira for helpful discussions during the
initial phase of this study.
This work was supported by CNPq, Fapemig, and the INCT (sistemas Complexos),
Brazil.

\end{document}